\documentclass[aip,jap,numerical,reprint,twoside]{revtex4-1}
\usepackage{graphicx}
\usepackage[OT4]{fontenc}
\begin{document}
\title{Plasmonic terahertz detectors based on a high-electron mobility GaAs/AlGaAs heterostructure}
\author{M.~Bia{\l}ek}
\email[]{marcin.bialek@fuw.edu.pl}
\author{A.~M.~Witowski}
\affiliation{Faculty of Physics, University of Warsaw, ul.\ Ho{\.z}a 69,
00-681 Warsaw, Poland}
\author{M.~Orlita}
\affiliation{Laboratoire National des Champs Magnétiques Intenses, CNRS-UJF-UPS-INSA, 25, avenue des Martyrs, 38042 Grenoble, France }
\author{M. Potemski}
\affiliation{Laboratoire National des Champs Magnétiques Intenses, CNRS-UJF-UPS-INSA, 25, avenue des Martyrs, 38042 Grenoble, France }
\author{M.~Czapkiewicz}
\affiliation{Institute of Physics, PAS, al.\ Lotnik\'{o}w 32/46, 02-668
Warsaw, Poland}
\author{J.~Wr\'{o}bel}
\affiliation{Institute of Physics, PAS, al.\ Lotnik\'{o}w 32/46, 02-668
Warsaw, Poland}
\affiliation{Faculty of Mathematics and Natural Sciences, Rzesz\'{o}w University, al.\ Rejtana 16A, 35-959 Rzesz\'{o}w, Poland}
\author{V.~Umansky}
\affiliation{Weizmann Institute of Science, Rehevot 76100, Israel}
\author{M.~Grynberg}
\affiliation{Faculty of Physics, University of Warsaw, ul.\ Ho{\.z}a 69,
00-681 Warsaw, Poland}
\author{J.~{\L}usakowski}
\affiliation{Faculty of Physics, University of Warsaw, ul.\ Ho{\.z}a 69,
00-681 Warsaw, Poland}
\begin{abstract}
In order to characterize magnetic-field ($B$) tunable THz plasmonic detectors, spectroscopy experiments were carried out at liquid helium temperatures and high magnetic fields on devices fabricated on a~high electron mobility GaAs/AlGaAs heterostructure. The samples were either gated (the gate of a~meander shape) or ungated. Spectra of a~photovoltage generated by THz radiation were obtained as a~function of $B$ at a~fixed THz excitation from a~THz laser or as a~function of THz photon frequency at a~fixed $B$ with a~Fourier spectrometer. In the first type of measurements, the wave vector of magnetoplasmons excited was defined by geometrical features of samples.
It was also found that the magnetoplasmon spectrum depended on the gate geometry which gives an additional parameter to control plasma excitations in THz detectors. Fourier spectra showed a~strong dependence of the cyclotron resonance amplitude on the conduction-band electron filling factor which was explained within a~model of the electron gas heating 
with the THz radiation. The study allows to define both the advantages and limitations of plasmonic devices based on high-mobility GaAs/AlGaAs heterostructures for THz detection at low temperatures and high magnetic fields.
\end{abstract}
\maketitle

\section{Introduction}
According to theoretical considerations, a~field-effect transistor (FET) can act as a detector of the electromagnetic radiation\cite{Dyakonov96}. Depending on material parameters, the length of the channel and the electron scattering time, the detection can be resonant or non-resonant. In the resonant case, plasma waves excited by the incident radiation reflect from the drain and source ends of the channel and create a~standing wave in an analogy with sound waves in an acoustic resonator. In theory, the frequency of the resonance depends on the electron concentration and can be tuned with the gate polarization. In the non-resonant case, the scattering time is too short to allow for generation of standing plasma waves and damped plasma oscillations develop only at the vicinity of the FET source. A~detailed discussion of these issues  has been recently presented in Ref.~\onlinecite{Knap10}.  

The theoretical proposal \cite{Dyakonov96} stimulated a~large number of experiments starting from the first demonstration of THz detection with a~FET a~decade ago \cite{Knap02nonresonant} (for a~review, see \onlinecite{KnapComtR}). Although the general concept of a~FET as a~THz detector was confirmed in a~large number of experiments, neither of known types of  FET showed, up to now, plasma resonances  at room-temperature. This concerns also a~recently investigated antenna-coupled graphene state of the art transistor\cite{Vicarelli12}. A~a~resonant detection was found only in some cases at liquid helium temperatures \cite{Knap02resonant, Elfatimy06, Dyer12}. On the other hand, a~strong non-resonant signal was observed on  transistors of different types, including Si metal-oxide-semiconductor FETs (Si-MOSFETs) \cite{Knap04, RabihAPLNEPSi}.

The current interest in room-temperature applications of FETs in THz detection seems to be split into two main research directions. On the one hand, technological and engineering efforts led to fabrication of multipixel cameras \cite{Pfeiffer12} based on a~non-resonant FET response. On the other, one tries to construct resonant antennas which -- due to their frequency characteristics -- could shrink a~non-resonant FET response into a~narrow-band one\cite{HustonPK, HustonJL}. 

The research on low-temperature detection with FETs is far less developed. There exists a~huge volume of theoretical and experimental results related to a~two-dimensional (2D) plasma investigation which started in 1977 with a~low-temperature Fourier spectroscopy on a~Si-MOSFET \cite{Allen77}. No doubt, from this perspective, different types of plasmon resonances in 2D systems seem to be perfectly well understood and analyzed. A~comprehensive review on this topic can be found in Ref.~\onlinecite{Kushwaha01}. However, developments of THz techniques, concepts and applications which have been created during last 20 years for room-temperature FET-based devices do not find their equally abundant counterpart in a low-temperature research. 

The present work aims to answer a~few questions which are related to the construction and performance of THz detectors fabricated for a~low-temperature THz magnetospectroscopy. The basic idea of the study was to investigate devices in which plasmon resonances could give a~strong and tunable signal. This led us to fabrication of FET-like devices with the design of the gate metallization formed as a~meander and an ungated reference sample of the same dimensions. Next, the detectors were tested in two different experiments: a~Fourier magnetospectroscopy and a~THz-laser magnetospectroscopy.
Let us note that the magnetic field is a~natural tuning parameter at low-temperature studies and is a~powerful tool allowing to interpret experimental results. In the Fourier spectroscopy, a~spectrum is acquired as a~function of the energy of photons at different magnetic fields. In the laser spectroscopy, the spectrum is measured as a~function of the magnetic field at a~constant wave length of incident THz photons.  
Testing the same detector in these two different experimental systems is not an usual approach since typically only one of these techniques is used. We show that carrying out these complementary experiments leads to a~much deeper characterization of the detector's performance. 

The paper is organized as the following. Section \ref{SamplesandExperiments} describes the samples and experimental techniques. Section \ref{Results} is split into two parts. The first one presents results of spectroscopic measurements obtained with monochromatic THz sources as a~function of the magnetic field. The second one is devoted to an~analysis of the Fourier spectroscopy measurements. The results are discussed in Section~\ref{Discussion}.

\section{Samples and Experiments}\label{SamplesandExperiments}
\begin{figure}
  \includegraphics[width=8.5cm]{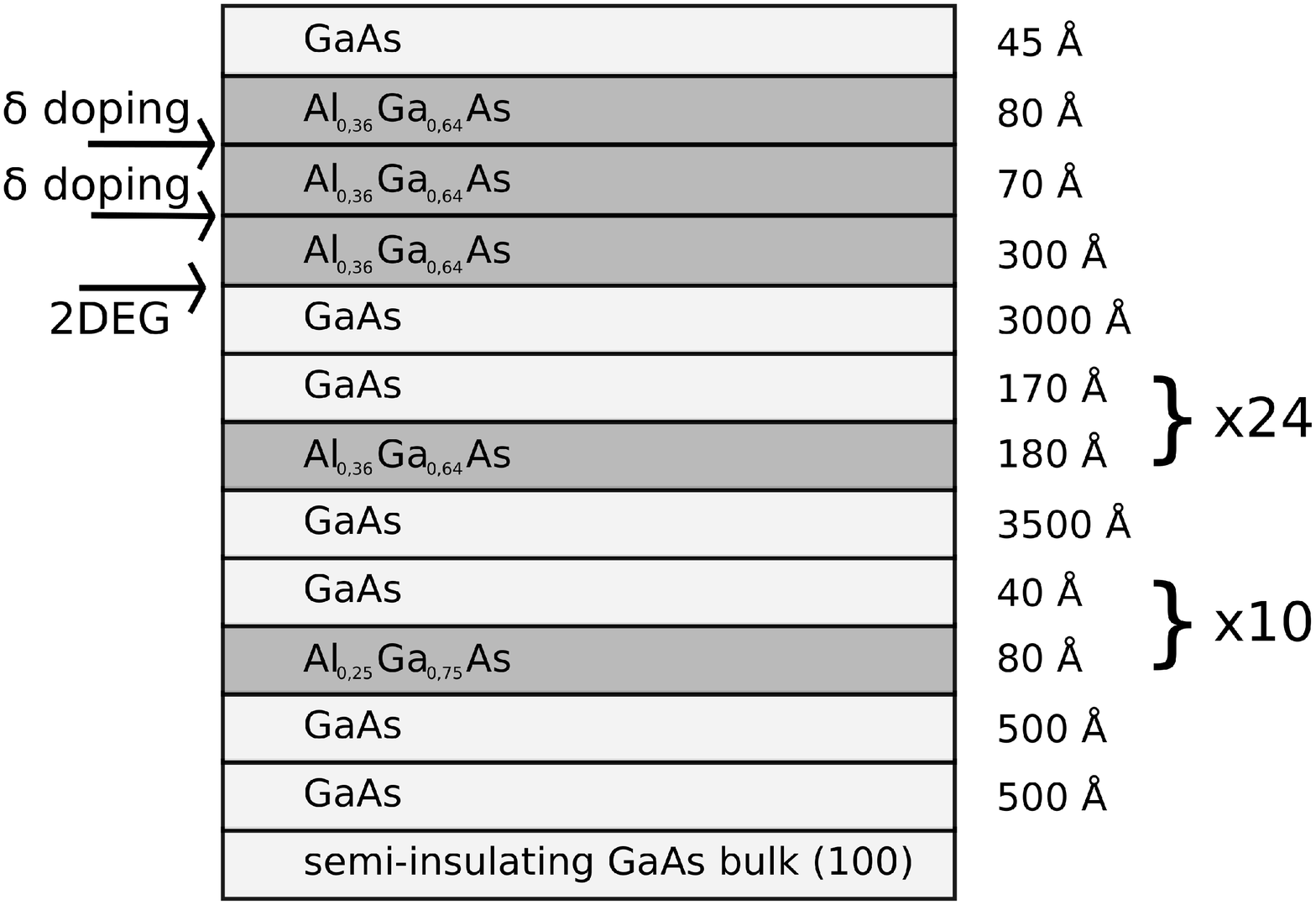}
 \caption{\label{Structure} A~scheme of the GaAs/AlGaAs heterostructure. The density of $\delta$ dopings is 7.5$\times$10$^{11}$~cm$^{-2}$.}
\end{figure}
\begin{figure}
 \includegraphics[width=8.5cm]{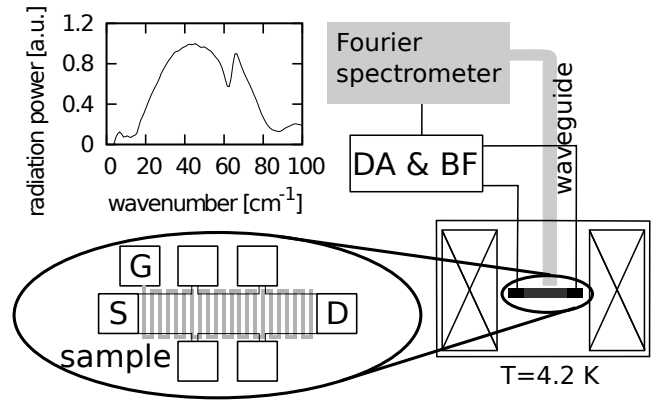}
 \caption{\label{lamp}Right: a~scheme of the experimental setup. Top inset: the spectrum of THz radiation incident on the sample. Bottom inset: the shape of a~meander-like gate on a~Hall bar sample; the photovoltage is measured between S and D contacts; G -- the gate electrode.}
\end{figure}
The samples were processed on a~high electron mobility GaAs/Al$_{0.36}$Ga$_{0.64}$As heterostructure with a~layer sequence shown in Fig.~\ref{Structure}. The electron mobility ($\mu$) in a~two-dimensional electron gas (2DEG) at 4~K was estimated by the Hall effect measurements to be about 6$\times$10$^5$~cm$^{2}$/Vs. The electron concentration ($N_s$) in ungated and gated samples was about 3.1$\times$10$^{11}$~cm$^{-2}$ and 2.65$\times$10$^{11}$~cm$^{-2}$, respectively. In gated samples, $N_s$ was typically about 15\% lower than in ungated ones. The thickness of the barrier $d$=48~nm will be used in the following paragraphs as one of parameters which determine the spectrum of plasmons. 

Two samples used for spectroscopic measurements were fabricated with an electron beam lithography on a~chip with dimensions of 4~mm~$\times~$4~mm. The mesas in the shape of Hall bars of dimensions of 1.300~mm~$\times$~0.065~mm were defined by a wet etching and were separated by about 1.5~mm (see Fig.~\ref{lamp} for a~scheme of the samples). Ohmic contacts were prepared by deposition of a~160~nm-thick layer of Au/Ge/Ni and heating the sample to 430$^o$C for a~few minutes. Connections of the four voltage contacts with the bar were done with 20~$\mu$m-thick channels.
The gate was evaporated on one of the mesas by deposition of a~15~nm-thick layer of Au/Pd alloy in a~meander form. The width of both gated and ungated stripes was equal to 2~$\mu$m. Let us note that due to the meander shape the period of the grid is equal to $a$~=~8~$\mu$m. This sample acted as a~grid-gated field-effect transistor. The chip was attached to a~metallic surface of a~14-pin dual-in-line support and both samples were bonded with Au wires.

The chip was placed in a~cryostat in a~center of a~superconducting coil and cooled down to 4.2~K with an exchange gas. The THz radiation was guided to the chip with an oversized waveguide made of a~stainless steel tube. The signal measured was a~photovoltage (PV) generated between the current-supplying contacts of a~Hall bar. During measurements, one of ohmic contacts was grounded, while Hall contacts were on a~floating potential. Measurements were carried out for both gated and ungated samples.

Two types of experiments were carried out. The first one was performed with a~monochromatic radiation from a~molecular THz laser pumped with a~CO$_2$ laser. Spectra were measured as a~function of the magnetic field. We used a~lock-in technique with the reference signal generated by a~mechanical chopper modulating on/off the laser beam. This type of measurements were carried out for a~few laser lines. In this setup we have also measured transport to determine the electron concentration and mobility.

The second one was a~Fourier spectroscopy experiment which gave the PV signal as a~function of the THz photon frequency at a~constant magnetic field. In this case, the measured signal was directed to a~differential amplifier with a~band-pass filter enabled and next to Bruker 113v Fourier spectrometer set to the spectral range of 20 to 80~cm$^{-1}$. The scheme of the experimental system as well as the radiation spectrum are shown in Fig.~\ref{lamp}. We collected Fourier spectra changing the magnetic field typically by 0.1~T. For gated samples we also measured spectra at a~constant magnetic field $B=3.0$~T and $5.2$~T, changing the gate polarization $U_G$ with 0.01~V steps. These values of the magnetic field were chosen because of a~strong PV signal and essentially different values of the conduction band electron filling factor $\nu$ at $U_G=0$~V. 

\section{Results} \label{Results}
\subsection{Laser spectroscopy}
\begin{figure}
 \includegraphics[width=8.5cm]{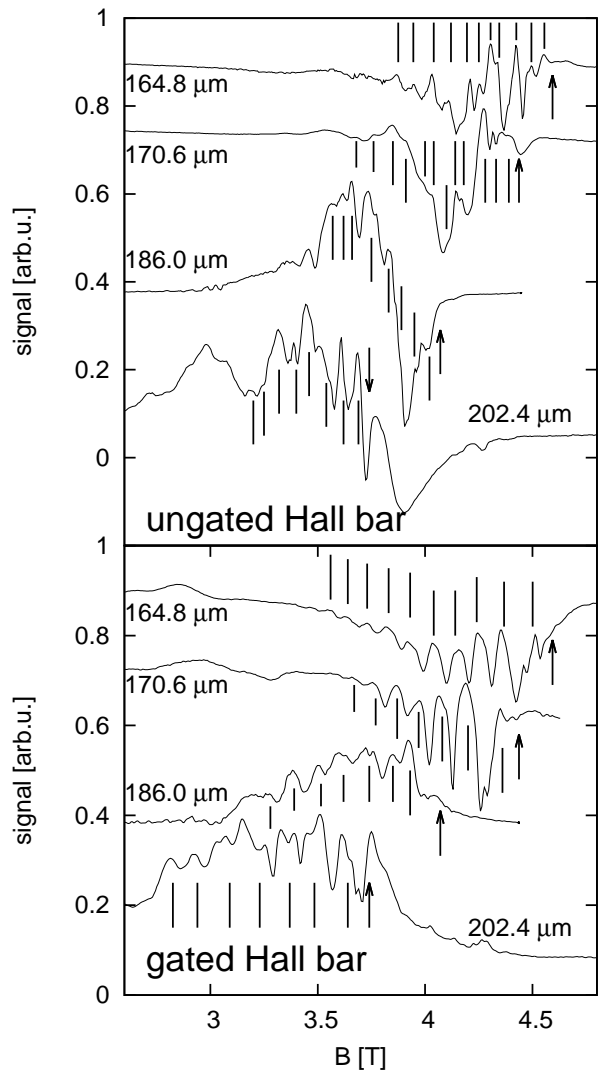}
\caption{\label{wyniki-laser}Normalized photovoltage spectra obtained at four THz laser lines (with the wavelength indicated in the figure) for gated (bottom) and ungated (top) samples at 4.2~K. Plasmon resonances are marked with bars. $B_{CR}$ is indicated with an arrow.}
\end{figure}
\begin{figure}
 \includegraphics[width=8.5cm]{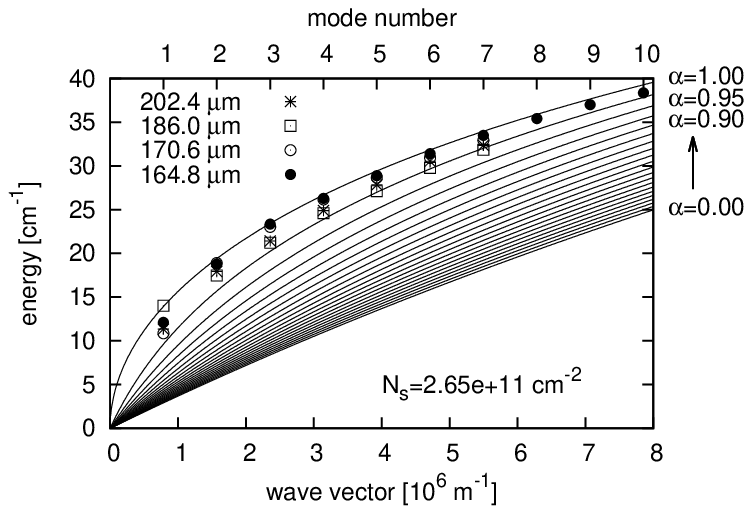}
\caption{\label{dyspersja-gated}Experimental (points) and theoretical (lines) dispersion of plasmon excited in the gated sample.}
\end{figure}
\begin{figure}
 \includegraphics[width=8.5cm]{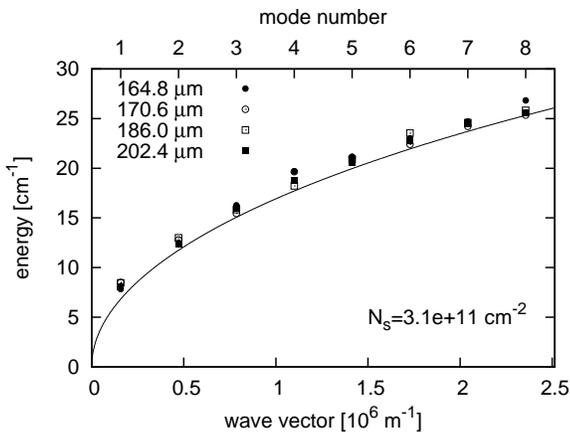}
\caption{\label{dyspersja-ungated}Experimental (points) and theoretical (lines) dispersion of plasmon excited in the ungated sample.}
\end{figure}

Let us discuss results of the molecular laser spectroscopy first. Typical spectra are shown in Fig.~\ref{wyniki-laser}. For both for gated and grid-gated sample, a~cyclotron resonance (CR) magnetic field $B_{CR}$ is preceded by a~series of maxima at lower magnetic fields. We interpret these peaks as magnetoplasmon resonances which are numbered with an integer $n=1,2,...$ counting from $B_{CR}$ toward lower fields. The frequency of magnetoplasmon follows the relation \cite{Theis77}:
\begin{equation}
 \omega^2=\omega_c^2+\omega_{p,n}^2,
 \label{mps}
\end{equation}
where $\omega$ is the magnetoplasmon frequency, $\omega_{p,n}$ is the frequency of $n$-th plasmon mode at zero magnetic field and $\omega_c=eB/m^*$ is the cyclotron  frequency; $e$ and $m^*$ are the electron charge and effective mass, respectively. The dispersion relation of plasmons for a~2D system at zero magnetic field is given by\cite{Stern67}:
\begin{equation}
 \omega_{p,n}=\sqrt{\frac{e^2N_s}{2m^{*}\epsilon_0}\frac{k_n}{\epsilon(k_n)}},
 \label{dispplasmon}
\end{equation}
where $N_s$ is the electron concentration, $k_n$ is the wavevector of the $n$-th plasmon mode and $\epsilon(k_n)$ is the effective dielectric function.

Depending on whether the sample is covered with a~metal gate or not, the effective dielectric function takes the form\cite{Eguiluz75}:
\begin{equation}
 \epsilon_{gated}(k)=\frac{1}{2}(\epsilon_{GaAs}+\epsilon_{GaAlAs}\coth(kd)),
 \label{dispgated}
\end{equation}
or\cite{Popov05}
\begin{equation}
 \epsilon_{ungated}(k)=\frac{1}{2}(\epsilon_{GaAs}+\epsilon_{GaAlAs}\frac{1+\epsilon_{GaAlAs}\tanh(kd)}{\epsilon_{GaAlAs}+\tanh(kd)}),
\label{dispungated}
 \end{equation}
where $d$~is the barrier thickness. In the case of grid-gated samples, a~weighted average of expressions given by Eqs.~\ref{dispgated} and~\ref{dispungated} is assumed\cite{KarolPhD}:
\begin{equation}
\epsilon(k)=\alpha\epsilon_{ungated}(\omega,k)+(1-\alpha)\epsilon_{gated}(\omega,k), 
\label{alpha}
\end{equation}
where $\alpha$~is a~parameter interpolating between two functions.

In Fig.~\ref{dyspersja-gated} we show experimentally determined frequencies of subsequent plasmon modes shifted to the zero magnetic field with Eq.~\ref{mps}. Theoretical plasmon dispersion curves, calculated at different values of the parameter $\alpha$ with Eqs.~\ref{dispgated}~--~\ref{alpha} are also presented. Calculations were carried out with the electron concentration $N_s = 2.65\times 10^{11}$~cm$^{-2}$ and $N_s = 3.1\times 10^{11}$~cm$^{-2}$ for gated and ungated sample, respectively. The effective mass $m^*= 0.0707m_e$ was determined from Fourier spectroscopy data described in the next section. Static dielectric constants, equal to 12.9 and 11.9 for GaAs and Al$_{0.36}$Ga$_{0.64}$As respectively, were used in calculations. With $k_n=2\pi n/a$ we get the best fit with the value of $\alpha\approx1.0$. 

In the case of the ungated sample to fit the experimental data with the Eq.~\ref{dispungated} we had to assume that observed magnetoplasmons were excited in the paths connecting the sample's channel with voltage side contacts. The width of these paths is $W_H=20$~$\mu$m. Due to the identical boundary conditions along both sides of the path, only wave vectors satisfying $k_n=(2n-1)\pi/W_H$, where $n=1,2,3,...$, modes can be excited under uniform spatial radiation distribution over the sample\cite{Alsmeier89, Kukushkin06, Mikhailov04}.
Under these assumptions we get a~good quantitative description of experimentally observed resonances shown in Fig.~\ref{dyspersja-ungated}. Generation of plasmons in samples without a~grid seems to contradict the general idea that a~grid coupler (i.e., a periodic structure) is necessary to couple incident photons with 2D plasma excitations, but the phenomenon is know for more than 20 years\cite{Alsmeier89, Vasiliadou93}. Let's note that the role of the grid is to diffract the radiation, so an 
electromagnetic wave acquires an in-plane wave vector component which allows to fulfill the momentum conservation law during excitation of a~plasmon with a~photon. Similarly, in grid-free samples, diffraction of photons on sample's or mesa's borders allows for plasmon generation.

\subsection{Fourier Spectroscopy}
\begin{figure}
 \includegraphics[angle=-90,width=8.5cm]{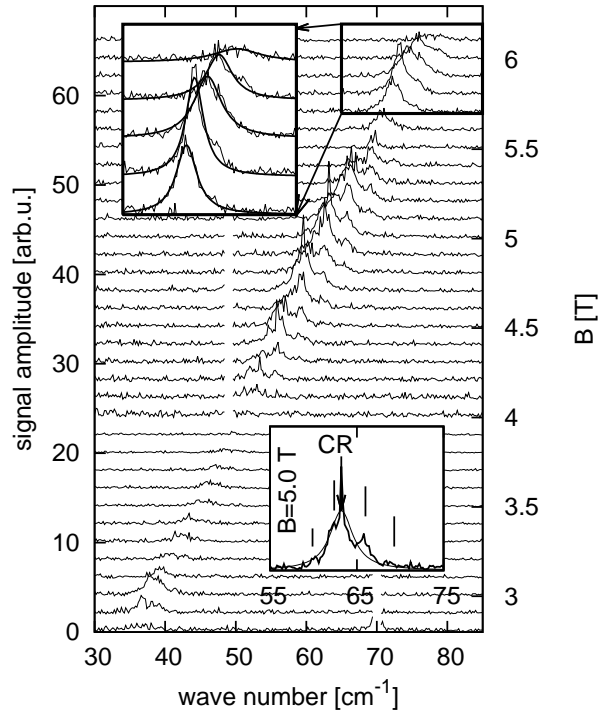} 
\caption{\label{H1-abs} Fourier spectra of PV measured  on the ungated sample at every 0.1~T between 2.8~T (bootom) to 6.1~T (top) at 4~K. Bars in the bottom inset show the maxima appearing due to interferences in the substrate.}
\end{figure}
\begin{figure}
 \includegraphics[angle=-90,width=8.5cm]{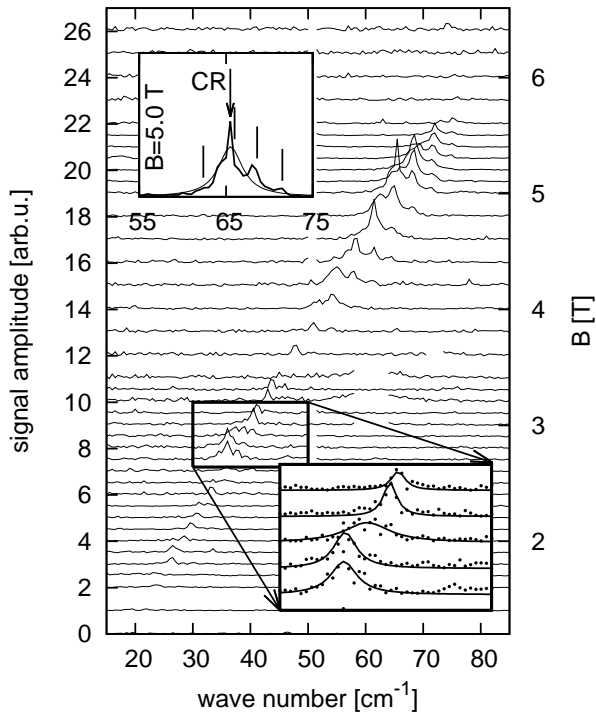} 
\caption{\label{H2-abs} Fourier spectra of PV measured  on the gated sample between 1.2~T (bootom) to 6.4~T (top) at 4~K. Bars in the top inset mark the maxima appearing due to substrate interference.}
\end{figure}


Fourier spectroscopy experiments were carried out for both gated and ungated sample. The resulting spectra collected at different magnetic fields for the ungated and gated samples are shown in Fig.~\ref{H1-abs} and Fig.~\ref{H2-abs}, respectively. Let us note that no PV signal was observed for radiation power below about 0.2~arb.u. of the intensity scale shown in the top inset in Fig.~\ref{lamp}. As one can notice, the spectra are dominated with a~symmetric peak of the full width at half maximum (FWHM) of about 5~cm$^{-1}$.
The peak is moving with the magnetic field towards higher frequencies and corresponds to the cyclotron resonance transition modified by magnetoplasmon excitations. The peak is broad enough to observe subsidiary maxima which appear with the period of (7.24$\pm$0.22)~cm$^{-1}$ (see the insets to Figs.~\ref{H1-abs} and \ref{H2-abs}). Since the period of these oscillations corresponds to interferences in the GaAs substrate, they are neglected in the~further analysis. For each spectrum, peaks were fitted with Lorentzian function to give estimations of their positions and amplitudes. The insets to Figs.~\ref{H1-abs} and~\ref{H2-abs} give examples of the fitting procedure.

\subsubsection{Frequency}
\begin{figure}
 \includegraphics[angle=-90,width=8.5cm]{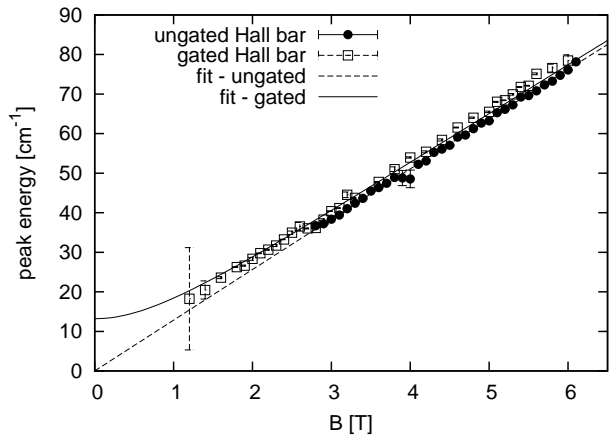}
 \caption{\label{frequency}Dependence of the magnetoplasmon-shifted cyclotron resonance peak frequency as a~function of the magnetic field.} 
\end{figure}
\begin{figure}
 \includegraphics[angle=-90,width=8.5cm]{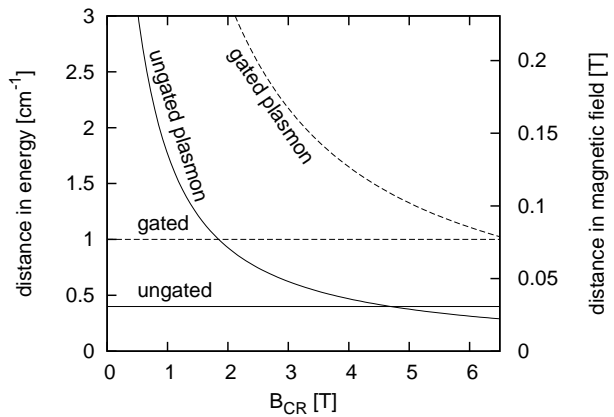}
 \caption{\label{resolution}Resolution limits of Fourier spectrometer for gated and ungated samples measurements and an evolution of a~difference in energy between plasmon mode and CR with the magnetic field for magnetoplasmon in both samples.}
\end{figure}

Data plotted in Fig.~\ref{frequency} show the position of the magnetoplasmon-shifted CR peak. We observe that the resonance frequency of the gated Hall bar is higher than that of the ungated one by a~few cm$^{-1}$. This results from the excitation of a~magnetoplasmon which blue-shifts the resonance frequency in the gated sample. Fitting Eq.~\ref{mps} to the gated sample data gives the plasmon frequency at zero magnetic field equal to (13.2$\pm$0.5)~cm$^{-1}$ which coincides with the frequency of $n=1$ mode  obtained from the THz laser experiment, shown in Fig.~\ref{dyspersja-gated}. Contrary to the THz laser spectroscopy results, we do not observe higher ($n > 1$) magnetoplasmon modes in the Fourier spectroscopy experiment which is caused by a~much smaller THz radiation power incident on the sample and its wide spectrum. 

We do not observe signatures of magnetoplasmon excitations in the ungated sample. This is related to the spectral resolution of the measurements which is discussed in Fig.~\ref{resolution}. The experiments were carried out with the resolution equal to 0.2~cm$^{-1}$ and 0.5~cm$^{-1}$ in the case of ungated and gated samples, respectively. We assume that the least distance between two maxima that can be resolved are twice these values which are represented with horizontal lines in Fig.~\ref{resolution}.
Curved lines show an~energy difference $\Delta E$ between the first magnetoplasmon mode and the cyclotron resonance: ($\sqrt{\omega_{p,1}^2+\omega_c^2} -\omega_c$) with $\omega_{p,1}=13$~cm$^{-1}$ and $\omega_{p,1}=7$~cm$^{-1}$ for gated and ungated sample, respectively (see Figs.~\ref{dyspersja-gated} and \ref{dyspersja-ungated}). One can notice, that with given experimental resolution, it is not possible to discriminate the plasmon from the CR for magnetic fields higher than about 4.5~T and 6.5~T for ungated 
and gated samples, respectively. This result means that at higher $B$~the peak observed in Fourier spectra coincides with the CR. The value of the cyclotron mass obtained from this data is equal to $0.0707m_e$. It is slightly higher than the value of $\approx$0.069$m_e$ expected in GaAs/AlGaAs heterojunction\cite{Batke88}. 
The right scale in Fig.~\ref{resolution} approximately recalculates $\Delta E$ into the magnetic field scale $\Delta E = e/m^* \Delta B$ which shows that the resolution of 0.005~T in the laser experimental system allows to discriminate between the CR and plasmon peaks even at high magnetic fields.

\subsubsection{Amplitude}\label{FSamp}
\begin{figure}
 \includegraphics[angle=-90,width=8.5cm]{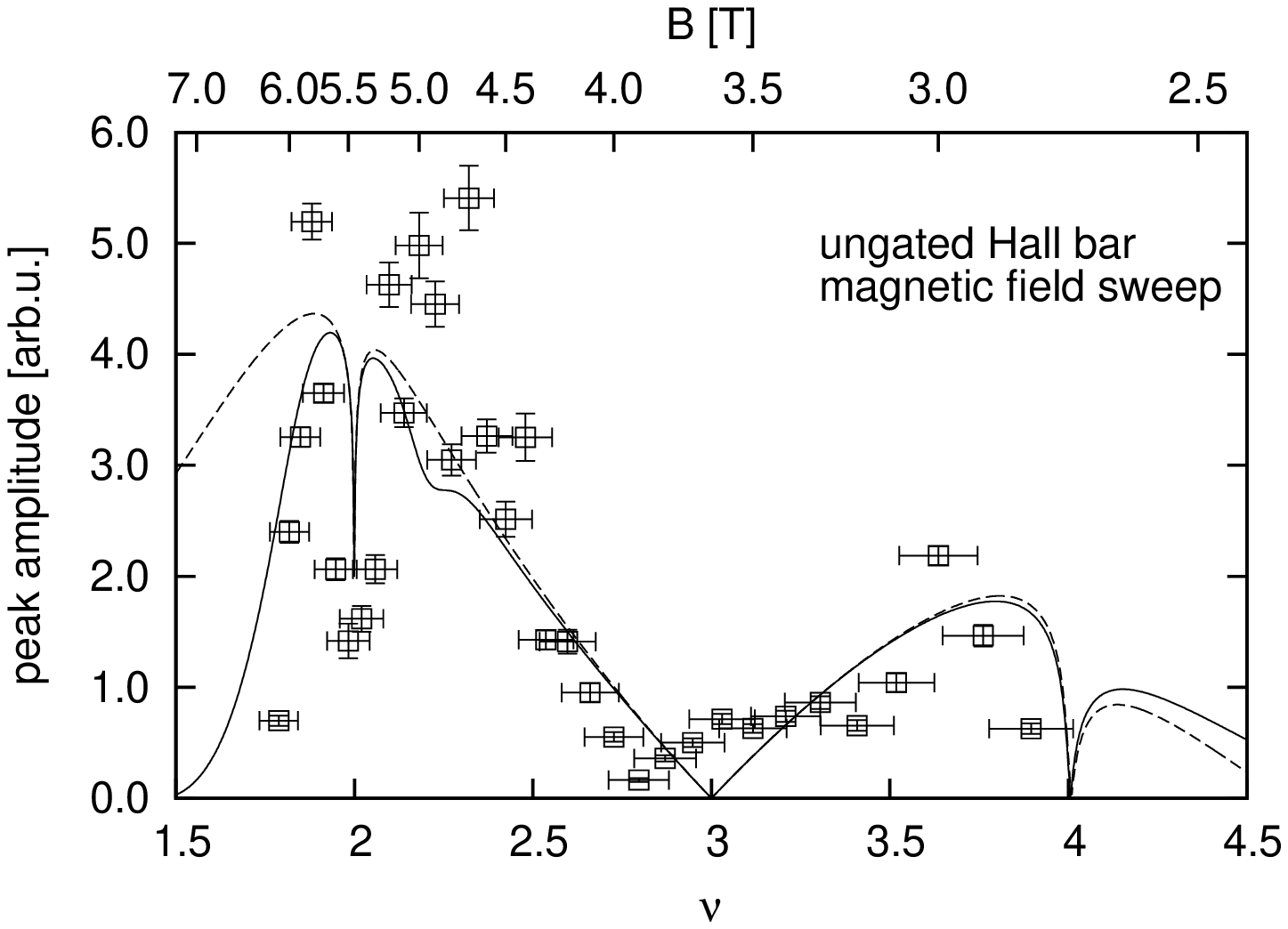}
 \includegraphics[angle=-90,width=8.5cm]{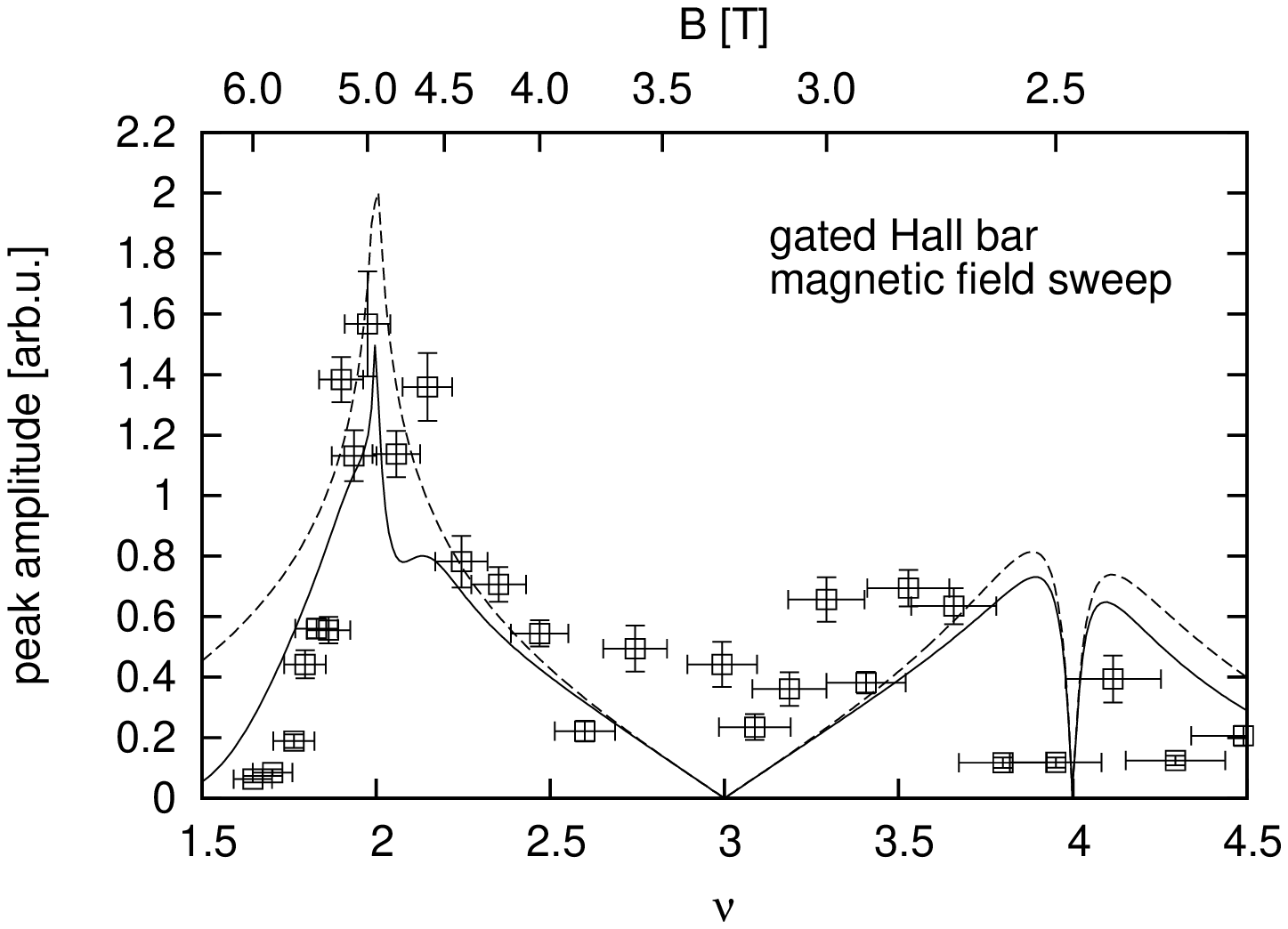}
 \includegraphics[angle=-90,width=8.5cm]{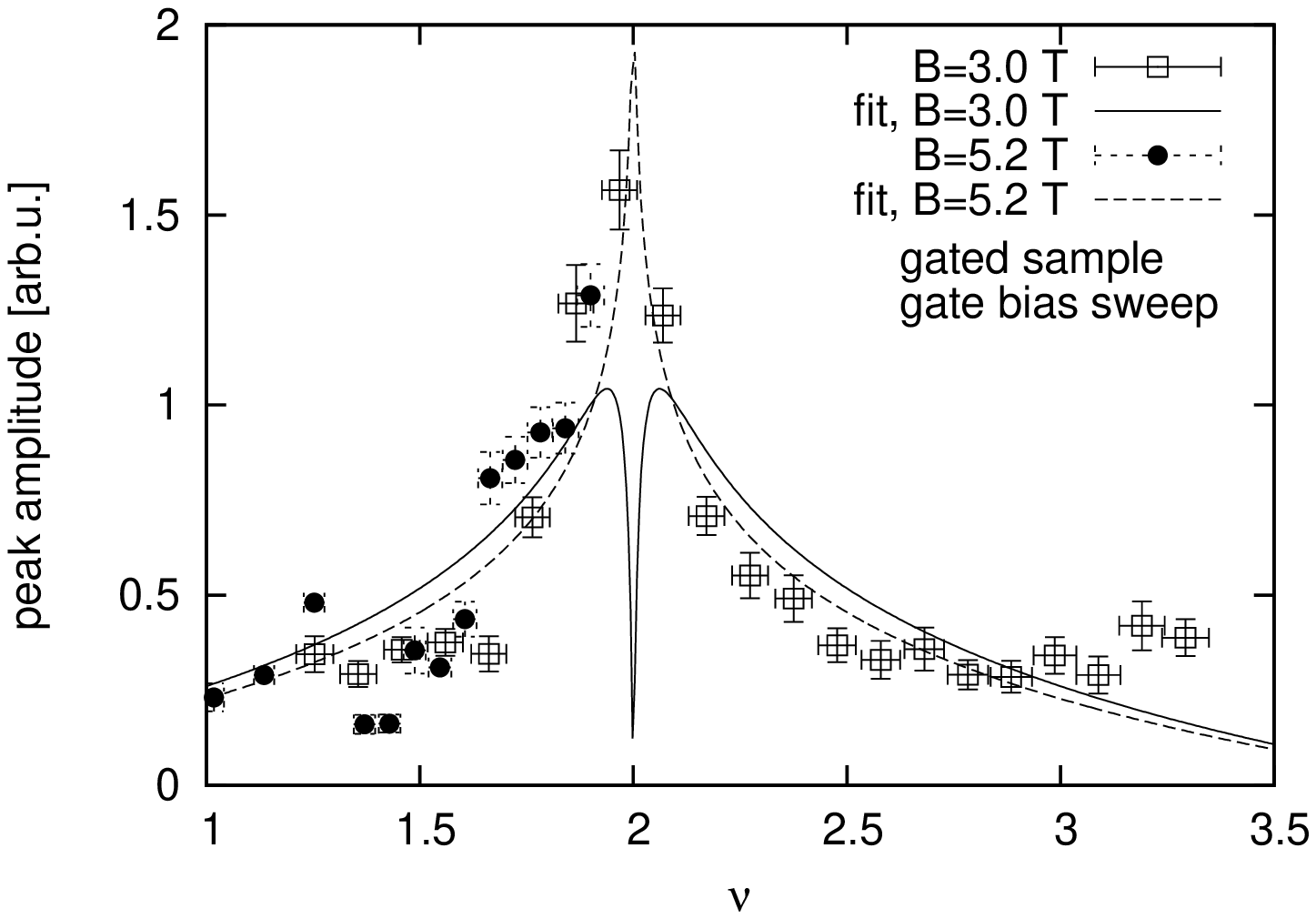}
 \caption{\label{fits}In the top and middle plots solid line show a~theoretically predicted PV including an~electron temperature dependence on the magnetic field. Dashed lines in these two plots show predictions with the constant electron temperature. In the bottom plot lines show predicted PV for a~constant $\Delta T_e$ in case of $U_G$ sweeps at a~constant $B$.}
\end{figure}

\begin{figure}
 \includegraphics[width=8.5cm]{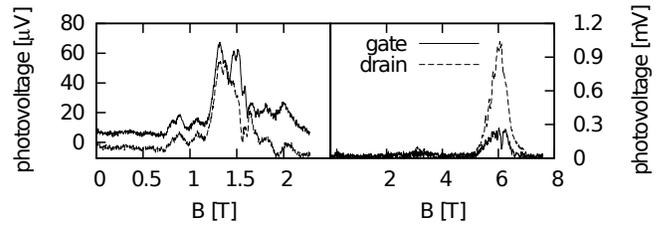}
 \caption{\label{PVUdsUg}Photovoltage measured on gate and drain contacts with monochromatic sources: 19.4~cm$^{-1}$ (left) and 84.2~cm$^{-1}$ (right).}
\end{figure}

The amplitude of the resonance present in Fourier spectra show a~non-monotonic dependence on the magnetic field. To analyze this dependence we refer to a~model described in Refs.~\onlinecite{Thiele89}. Its essence lies in an assumption that absorption of a~THz radiation changes the electron distribution function. This can be approximately described as a~change of the electron temperature $T_e$, which becomes higher than the lattice temperature $T_L$. In a~gated sample, there is a~strict correspondence between the gate potential and the electron distribution function (it just reflects the fact that the electron concentration can be changed with the gate polarization). For this reason, changes in the electron temperature can be observed as changes of the gate -- to -- channel potential, as it was done in Ref.~\onlinecite{Thiele89} where the gate -- to -- channel photovoltage was investigated. In our case, typically a~drain -- to -- source photovoltage was measured (see the inset to Fig.~\ref{lamp} to recall 
the names 
of the sample's contacts) which was proved to be essentially identical to that of a~gate -- to -- source, as Fig.~\ref{PVUdsUg} shows. This similarity must exists since in gated samples gate -- to -- source and drain -- to -- source potentials are dependent under the open drain conditions, as it is the case of a~PV measurements. To explain the origin of a~PV signal in an ungated sample we note that an asymmetry must be present in the sample which allows for a~charge separation and generation of a~PV under the incidence of THz radiation. Since the process of charge separation depends on the electron diffusion, it is influenced by the electron distribution function, hence $T_e$. 

The model proposed in Ref.~\onlinecite{Thiele89} allows to estimate the value of the photovoltage for magnetic fields corresponding to electron filling factors close to even values, $\overline\nu = 2, 4, ...$:
\begin{eqnarray}
\label{wzor}
\nonumber
\Delta V(\nu)=A\{T_e\mathrm{arcsinh}(\frac{\nu-\overline\nu}{4}\exp[\frac{\hbar\omega_c}{2kT_e}])\\
-T_L\mathrm{arcsinh}(\frac{\nu-\overline\nu}{4}\exp[\frac{\hbar\omega_c}{2kT_L}]
)\},
\end{eqnarray}
where $A$ is a~fitting parameter and $T_L$ = 4.2~K. For the gated sample, the filling factor $\nu$~was calculated as a~function of $B$~and the gate voltage $U_G$:
\begin{equation}
\nu=\frac{h}{e}N_s\frac{U_G-U_{th}}{U_{th}}\frac{1}{B},
\label{ff}
\end{equation}
where the fraction ($U_G-U_{th})/U_{th}$, with $U_{th}$ denoting a~threshold voltage, gives a~fraction of free electrons with respect to their number at zero gate polarization. The value of $U_{th}$=-0.323$\pm$0.006~V was obtained from a~dependence of the electron concentration on the $U_G$ determined by an analysis of the period of SdH oscillations; for the ungated sample, this fraction was $1$. The correspondence between $\nu$~and $B$~allows to express $\omega_c$ as a~function of $\nu$~in Eq.~\ref{wzor}. In the experiment, only the amplitude of the photovoltage was measured. For this reason, the modulus of the function~\ref{wzor} was fitted to the data. Our measurements were done for magnetic fields corresponding, approximately, to $1.5 < \nu < 4.5$. For this reason, the fitting function was composed of a~sum of two contributions, one corresponding to $\overline\nu = 2$ and the other -- to $\overline\nu = 4$. 

We estimated $T_e$ in two ways. First, we assumed that $T_e$ is constant, i.e., it does not depend on the magnetic field. This is an approximation since heating of the electron gas is the most efficient with photons of the energy close to that of the CR and the spectral intensity of the THz source in the Fourier spectrometer is not constant. That is why, we assumed that $T_e(B) = T_L +\Delta T l(B)$, where $l(B)$ is a~normalized power of the THz source, shown in Fig.~\ref{lamp}, with the argument $B$~determined according to $E = \hbar \omega_c$, where $E$~is the photon energy. This model gives the maximal heating efficiency at the magnetic field equal to about 3.7~T which corresponds to the CR at $E=45$~cm$^{-1}$.

Finally there were two fit parameters only: $\Delta T_e$ -- the difference between temperatures of electrons and the lattice (4.2~K) and $A$ -- a~scaling factor for the function (\ref{wzor}). Results of both fitting procedures are shown in Fig.~\ref{fits}. One can notice that the model describes the experimental data with a~reasonable accuracy. Also, a~better fitting results are obtained with an assumption of a~$B$-dependent electron temperature (dashed lines) than with a~constant one (solid lines). Fitted values of $\Delta T$ are equal to 21.0~K and 3.0~K for the ungated and gated samples, respectively. The same value of $A$~was used for both samples. Taking into account that the PV signal was observed for the source intensity higher than 0.2 of the maximum value, the electron temperature changes between (4.2 + 0.2*21.0)~K = 8.4~K to 25.5~K for the ungated sample and between 4.8~K and 7.2~K for the gated one. 

An analysis of Eq.~\ref{wzor} shows that a~sharp feature in $\Delta V$ should be expected at $\nu= 2, 4, ...$. In fact, such abrupt changes are clearly visible in the data presented in Fig.~\ref{fits}. Since transport measurements were not done during the Fourier spectroscopy measurements, we estimated the value of the filling factor by assuming that $\nu = 2$ corresponds to $B$ = 5.6~T and 5.0~T in the case of ungated and gated samples, respectively. This leads, correspondingly, to $N_s = 2.65\times10^{11}$~cm$^{-2}$ and $N_s=2.4\times10^{11}$~cm$^{-2}$. These values are by about 10-15\% less than obtained at the THz laser spectroscopy and magnetotransport measurements ($3.1\times10^{11}$~cm$^{-2}$ and $2.65\times10^{11}$~cm$^{-2}$) which can result from the dependence of $N_s$ on a~particular cooling cycles in these two separate experiments. 

\section{Discussion} \label{Discussion}

The most evident application of the investigated samples as low-temperature tunable THz detectors results from the Fourier spectroscopy measurements. Neglecting the interference in the substrate (which can be eliminated by wedging the sample), a~linear dependence of the resonance energy on the magnetic field together with a~Lorenzian shape of the peak, shown in Figs.~\ref{H1-abs} and \ref{frequency}, is suitable for application at high enough $B$. At lower fields one can observe a~deviation from the linearity related to excitation of magnetoplasmons.
Thus, designing a~detector one should be aware of magnetoplasmons that can be generated by the radiation. According to our analysis, it is possible to estimate the range of magnetic fields at which the separation of the magnetoplasmon from the CR position is lower than the experimental resolution (see Fig.~\ref{resolution}). On the other hand, oscillations of the resonance intensity with the electron filling factor, observed also by other groups\cite{Seidenbusch87, Thiele89, Wilson79, Heitmann86, Batke88, Tsujino98}, can limit the applicability of the detector since one can not trace the investigated signal in a~broad range of $B$~with the same sensitivity. 

It is not obvious to determine by theoretical considerations the wave vector of the fundamental plasmon mode in a~given sample. The thing is that in a~real detector one always finds some metallization pads or signal-conducting paths which can trigger plasmon excitation with a~wave vector determined by a~geometrical detail of the structure. This is clearly seen in the investigated samples where under the same conditions of laser excitations, plasmons were generated according to the geometry of the grid-gate or side signal paths (in this case, paths leading to Hall voltage pads).
The energy difference between the fundamental modes in these two cases is equal to about 7~cm$^{-1}$ which is far more than the spectral resolution of measurements. Testing the same detectors at much lower frequencies, in the range 0.1 -- 0.6~THz, we found that we were able to excite plasmons with the wave vector of the fundamental mode equal to the width of the mesa, i.e., 65~$\mu$m.\cite{Bialek12} Taking all these observations into account we conclude that the spectrum of plasmons  generated  depends both on the range of the radiation frequency used and geometrical details of a~given detector design. This variety of possible plasmons generated in one sample can be considered as discouraging, but we rather see in this observation a~chance to control plasmon resonances by a~proper design of the detector geometry.
For instance, if in the ungated sample investigated the fundamental mode of the plasmons corresponds to the width of a~side path, one could propose a~device with tens of such paths positioned at the circumference of the mesa. Such a~design should lead to a~strong amplification of the plasmonic response. 

Summing up, in a~Fourier spectroscopy experiments, the detectors investigated showed a~clear and narrow spectral response but the amplitude of the signal strongly depends on the magnetic field. The narrow spectral width results from the fact that only one magnetoplasmon mode was generated in the Fourier experiment. Since the variations of the amplitude of the detection signal result from basic physical properties of a~2DEG, we consider this drawback as hard to eliminate. In a~THz laser spectroscopy, the signal is spectrally much broader and dominated by a~number of plasmon resonances triggered according to the geometrical details of the detector design and the range of THz frequency radiation used. We suggest that the geometry of a~properly designed detector could stimulate generation of only one plasmon mode which could lead to a~narrow response in THz laser spectroscopy experiments. 

\section{Conclusions}
Gated and ungated Hall bars were used in a~THz magnetospectroscopy experiments. Excitation with a~THz laser allowed to determine dispersion relations of plasmons. In the case of the gated sample, magnetoplasmons were generated with the wave vector of the fundamental mode equal to the period of the grid. For the ungated sample, it was determined by details of the geometry of the mesa. Fourier spectroscopy showed that the amplitude of the magnetoplasmon resonance was an oscillating function of the electron filling factor which can be described by a~model assuming an increase of the electron gas temperature. These two types of experiment gave consistent spectral results.

\section{Acknowledgments}
This work was partially supported by Foundation for Polish Science grant POMOST/2010-1/8. We are grateful to dr K.~Fronc for the help in samples processing.

\providecommand{\noopsort}[1]{}\providecommand{\singleletter}[1]{#1}%

\end{document}